\documentclass{appolb}
\usepackage{epsfig}

\usepackage{graphics}
\usepackage{amssymb}
\usepackage{amsmath}

\newcommand{\rf}[1]{(\ref{#1})}
\newcommand{\beq}{\begin{equation}}

\newcommand{\eeq}{\end{equation}}
\newcommand{\bea}{\begin{eqnarray}}
\newcommand{\eea}{\end{eqnarray}}

\renewcommand{\e}{\mbox{e}}
\renewcommand{\d}{\mbox{d}}

\newcommand{\G}{\Gamma}

\renewcommand{\b}{\beta}

\newcommand{\m}{\mu}

\newcommand{\sg}{\sigma}

\newcommand{\ra}{\rangle}
\newcommand{\la}{\langle}
\newcommand{\prt}{\partial}
\newcommand{\mi}{\!-\!}
\newcommand{\equ}{\!=\!}
\newcommand{\pl}{\!+\!}

\newcommand{\cO}{{\cal O}}

\newcommand{\tP}{{\tilde{P}}}
\newcommand{\tG}{{\tilde{G}}}

\newcommand{\hn}{{\hat{n}}}

\newcommand{\bx}{{\bar{x}}}
\newcommand{\bt}{{\bar{t}}}

\newcommand{\sm}{\sqrt{\m}}

\newcommand{\cm}{{\rm cardy}}
\newcommand{\zz}{{\rm zz}}

\newcommand{\rara}{\ra\!\ra}
\newcommand{\vv}{|}

\begin{document}
\title{ZZ branes from a worldsheet perspective.
\thanks{Based on a presention  at the RMT workshop in Krakow, May 3-5, 2007}%
}
\author{Jan Ambj\o rn$\,^{a,b}$ and Jens Anders Gesser$\,^a$
\address{$^a$~The Niels Bohr Institute, Copenhagen University\\
Blegdamsvej 17, DK-2100 Copenhagen \O , Denmark.\\}
\address{$^b$~Institute for Theoretical Physics, Utrecht University, \\
Leuvenlaan 4, NL-3584 CE Utrecht, The Netherlands.}
}
\maketitle
\begin{abstract}
We show how non-compact space-time (ZZ branes) emerges as a limit of
compact space-time (FZZT branes) for specific ratios between the
square of the boundary cosmological constant and the bulk
cosmological constant in the (2,2m - 1) minimal model coupled to
two-dimensional euclidean quantum gravity. Furthermore, we show that
the principal (r,s) ZZ brane can be viewed as the basic (1,1) ZZ
boundary state tensored with a (r,s) Cardy boundary state for a
general (p,q) minimal model coupled to two-dimensional quantum
gravity. In this sense there exists only one ZZ boundary state, the
basic (1,1) boundary state.

\end{abstract}

\section{Introduction}

Two-dimensional Euclidean quantum gravity serves as
a good laboratory for the study of potential theories
of quantum gravity in higher dimensions. Although it
contains no dynamical gravitons and does not face the problem
of being non-renormalizable, it can address many of the
other conceptional questions which confronts a {\it quantum field theory}
of gravity. How does one define the concept of distance in a theory
where one is instructed to integrate over all geometries, how
does one define the concept of correlation functions when one
couples matter to gravity and the resulting theory is
supposed to be diffeomorphism invariant? These are just two of
many questions which can be addressed successfully and which
are as difficult to answer in two dimensions as in higher dimensions.
In addition two-dimensional quantum gravity coupled to a minimal
conformal field theory is nothing but a so-called
non-critical string theory and
serves as a good laboratory for the study
of non-perturbative effects in string theory.

The quantization of 2d gravity was first carried out for compact
two-dimensional geometries using matrix models, combinatorial
methods and methods from conformal field theory. Later on
Zamolodchikov, Zamolodchikov and Fateev and also Teschner (FZZT)
used Liouville quantum field theory to quantize the disk
geometry\cite{fzz}, thereby reproducing results already obtained
from matrix models. Then the Zamolodchikovs (ZZ) turned their
attention to a previously unadressed question crucial to quantum
gravity, namely how to quantize non-compact 2d Euclidean geometries
\cite{zz}. They asked if the quantization of the disk geometry could
be generalized to the Lobachevskiy plane, also known as Euclidean
AdS$_{2}$ or the pseudosphere. The pseudosphere is a non-compact
space with no genuine boundary. However, one has to impose suitable
boundary conditions at infinity in order to obtain a conformal field
theory. The crucial difference compared to the quantization of the
compact disk is that one can invoke the assumption of factorization
when discussing the correlator of two operators. One assumes that
\beq\label{0.0} \la \cO_1 (x) \cO_2(y)\ra \to \la \cO_1(x)\ra \la
\cO_2(y)\ra \eeq when the {\it geodesic distance on the
pseudosphere} between $x$ and $y$ goes to infinity. This additional
requirement results in a number of self-consistent boundary
conditions at infinity compatible with the conformal invariance of
quantum Liouville theory.

It is the purpose of this article to address the question of
quantizing non-compact 2D Euclidean geometries using a different
approach than the Zamolodchikovs. However, our results will relate
to the random geometries obtained by the Zamolodchikovs.

In modern string terminology boundary conditions are almost
synonymous to ``branes'' and in this spirit the conventional
partition function for the disk and the partition function for the
pseudosphere were reinterpretated in non-critical string theory as
FZZT and ZZ branes, respectively. In this context it was first
noticed that there is an intriguing relationship between the FZZT
and the ZZ branes \cite{martinec,sei1,sei2,martinec2}, as well as
between the analytical continuation of the disk amplitude to the
complex plane and the space of conformal invariant boundary
conditions one can impose.

It is the objective of this article to analyze these relations from
a worldsheet perspective.

\section{From compact to non-compact geometry}

The disk and cylinder amplitudes for generic values of the coupling
constants in minimal string theory were first calculated using
matrix model techniques. In order to compare with continuum
calculations performed in the context of Liouville theory, it is
necessary to work in the so-called conformal background
\cite{staudacher}. In the following we will, for simplicity,
concentrate on the disk and the cylinder amplitudes in the $(2,2m
\mi 1)$ minimal conformal field theories coupled to 2d quantum
gravity. In the conformal background the disk amplitude is given by:
\beq\label{1} 
w_\m(x) = (-1)^{m}\hat{P}_m(x,\sm)\sqrt{x+\sm} =
(-1)^{m}\left(\sm\right)^{(2m-1)/2} P_m(t)\sqrt{t+1},
\eeq 
where $t=x/\sm$ and where \cite{sei1,staudacher} 
\beq\label{2}
P^2_m(t)\;(t+1) = 2^{2-2m}(T_{2m-1}(t)+1), 
\eeq 
$T_p(t)$ being the
first kind of Chebyshev polynomial of degree $p$. In eq.\ \rf{1} $x$
denotes the boundary cosmological coupling constant and $\m$ the
bulk cosmological coupling constant, the theory viewed as 2d quantum
gravity coupled to the $(2,2m \mi 1)$ minimal CFT. The zeros of the
polynomial $P_m(t)$ are all located on the real axis between $\mi 1$
and $1$ and more explicitly we can write: 
\beq\label{2a} 
P_m(t) =\prod_{n=1}^{m-1} (t-t_n),~~~~~ t_n =
-\cos\left(\frac{2n\pi}{2m-1}\right),~~~~1\le n \le m-1. 
\eeq

The zeros of $P_m(t)$ can be associated with the $m\mi 1$ principal
ZZ branes in the notation of \cite{sei1}. In order to {\it
understand this}, i.e.\ in order to understand why the special
values $t_n$ (and only these values) of the boundary cosmological
constant are related to non-compact worldsheet geometries, it is
useful to invoke the so-called loop-loop propagator $G_\m(x,y;d)$
\cite{kawai1,kawai2,gk,watabiki,aw_pq}. It describes the amplitude
of an ``exit'' loop with boundary cosmological constant $y$  to be
separated a distance $d$ from  an ``entrance'' loop with boundary
cosmological constant $x$ (the entrance loop conventionally assumed
to have one marked point). $G_\m(x,y;d)$ satisfies the following
equation: 
\beq\label{5} 
\frac{\prt }{\prt d}\; G_\m (x,y;d) = -
\frac{\prt }{\prt x} \; w_\m(x) G_\m (x,y;d), 
\eeq 
with the following solution: 
\beq\label{6} 
G_\m (x,y;d) =\frac{w_\m(\bx(d))}{w_\m(x)} \; \frac{1}{\bx(d)+y},
~~~~~~~d=\int_{\bx(d)}^x \frac{\d x'}{w_\m(x')}, 
\eeq 
where $\bx(d)$ is called the running boundary coupling constant.

For the $(2,2m\mi 1)$ minimal model coupled to 2d gravity \rf{6}
reads: 
\beq\label{6a} 
G_\m(t,t';d) \propto \frac{1}{\sm}\;\;\frac{1}{ \bt(d) + t'}\;\;
\frac{\sqrt{1+\bt(d)}\;\prod_{n=1}^{m-1} \;(\bt(d) -
t_n)}{\sqrt{1+t}\; \prod_{n=1}^{m-1} \;(t - t_n)} 
\eeq 
where we use the notation of \rf{1}, 
i.e.\ $t=x/\sm$, $t'=y/\sm$ and $\bt(d) =\bx(d)/\sm$.
For $m =2$, i.e.\ pure gravity
$d$ measures the geodesic distance. For $m>2$ this is not true.
Rather, it is a distance measured in terms of matter excitations.
This is explicit by construction in some models of quantum gravity
with matter, for instance the Ising model and the $c\equ \mi 2$
model formulated as an $O(-2)$ model \cite{akw,aajk}. However, we
can still use $d$ as a measure of distance and we will do so in the
following. When $d \to \infty $ it follows from \rf{6} that the
running boundary coupling constant $\bt(d)$ converges to one of the
zeros of the polynomial $P_m(t)$, i.e.\ 
\beq\label{6b} 
\bt(d) \xrightarrow[d\to \infty]{~} t_k,~~~t_k =
-\cos\left(\frac{2k\pi}{2m-1}\right). 
\eeq

The cylinderamplitude (\ref{6a}) vanishes for generic values of $t'$
in the limit $d \to \infty$. However, as shown in \cite{aagk} we
have a unique situation when we choose $t' = -t_k$ since in this
case the term $1/(\bt(d)+t')$ in \rf{6a} becomes singular for $d \to
\infty$. After some algebra we obtain the following expression:
\bea\label{6c}
G_\m (t, t'=-t_k,d\to \infty) &\propto&
\frac{1}{\sm}\;\frac{1}{\sqrt{1+t}} \sum_{n=1}^{m-1}
(-1)^n \sin\left(\frac{2n\pi}{2m-1} \right) \\
&& \left[ \frac{1}{\sqrt{1+t}+\sqrt{1+t_n}}-
\frac{1}{\sqrt{1+t}-\sqrt{1+t_n}}\right]. \nonumber 
\eea
Notice, $G_\m(t, t'=-t_k,d\to \infty)$ is independent of which zero $t_k$
the running boundary coupling constant approaches in the limit $d
\to \infty$, apart from an overall constant of proportionality.

Formula \rf{6c} describes an AdS-like non-compact space with
cosmological constant $\m$ and with one compact boundary with
boundary cosmological constant $x$ as explained in \cite{aagk} in
the case of pure gravity. In the last section we will comment on the
fact that we have to set $t' = -t_k$ in order to generate an
AdS-like non-compact space in the limit $d \to \infty$ and that
$t_{k}$ serves as an attractive fixed point for the running boundary
coupling constant. Now, we will explain how the cylinder amplitude
(\ref{6c}) is related to the conventional FZZT--ZZ cylinder
amplitude in the Liouville approach to quantum gravity.

\section{The cylinder amplitudes}

Like the disk amplitude \rf{1}, the cylinder amplitude in the
$(2,2m\mi 1)$ minimal CFT coupled to 2d quantum gravity was first
calculated using the one-matrix model. Quite remarkable it was found
to be universal, i.e.\ the same in all the $(2,2m\mi 1)$ minimal
models coupled to quantum gravity \cite{staudacher,ajm}:
\beq
Z_\m(t_1,t_2) = - \log \left[\Big(\sqrt{t_1+1}+\sqrt{t_2+1}\Big)^2
\sm \, a\right],
\label{8a}
\eeq
where $a$ is a (lattice) cut-off.

The amplitude $Z_\m(t_1,t_2)$ is only one of many cylinder
amplitudes which in principle exist when we consider a $(2,2m\mi 1)$
minimal conformal field theory coupled to 2d gravity. If we consider
the cylinder amplitude of the $(2,2m\mi 1)$ minimal conformal field
theory before coupling to gravity we have available $m\mi 1$ Cardy
boundary states $|r\ra_{\mathrm{Cardy}}$, $r\equ 1,\ldots,m\mi 1$,
on each of the boundaries, and a corresponding cylinder amplitude
for each pair of Cardy boundary states \cite{bppz}:
\beq\label{9}
Z_{matter}(r,s;q) = \sqrt{2}\; b \sum_{l=1}^{m-1} (-1)^{r+s+m+l+1}
 \frac{\sin (\pi r l b^2)
\sin (\pi s l b^2)}{\sin(\pi  l b^2)}  \, \chi_l (q),
\eeq
where
\beq\label{9a} b=\sqrt{\frac{2}{2m-1}}
\eeq
and where we consider a
cylinder with a circumference of $2\pi$ and length $\pi \tau$ in the
closed string channel. The generic non-degenerate Virasoro character
$\chi_p(q)$ is
\beq\label{10}
\chi_p(q) =\frac{q^{p^2}}{\eta(q)},~~~~q=\e^{-2\pi \tau},
\eeq
where $\eta(q)$ is the Dedekind function.
However, the degenerate Virasoro character
$\chi_l(q)$ in eq.\ \rf{9} is given by \cite{book}:
\beq\label{11}
\chi_l(q)= \frac{1}{\eta(q)}\; \sum_{n \in \;\mathbb{Z}} \left(
q^{(2n/b+1/2(1/b -l\, b))^2} - q^{(2n/b+1/2(1/b+l\, b))^2}\right).
\eeq

In order to couple  the cylinder amplitude in eq.\ \rf{9} to 2d
quantum gravity one has, in the conformal gauge, to multiply
$Z_{mat}(r,s;q)$ by a contribution $Z_{ghost}(q)$ obtained by
integrating over the ghost field, as well as by a contribution
$Z_{Liouv}(t_1,t_2;q)$ obtained by integrating over the Liouville
field. Explicitly we have
\beq\label{12} 
Z_{ghost}(q)=
\eta^2(q),~~~~Z_{Liouv}(t_1,t_2;q)= \int_0^\infty \d P
\;\bar{\Psi}_{\sg_{1}}(P) \Psi_{\sg_{2}}(P) \chi_P(q),
\eeq
where $\Psi_\sg(P)$ is the FZZT boundary wave function \cite{fzz}, such
that
\beq\label{13} 
\bar{\Psi}_{\sg_{1}}(P) \Psi_{\sg_{2}}(P) =
\frac{4 \pi^2 \cos(2\pi P \sg_{1}) \cos(2\pi P \sg_{2}) }{\sinh
(2\pi P/b ) \sinh (2\pi P b ) },
\eeq
and where $\sg$ is related to
the boundary cosmological constant by
\beq\label{14}
\frac{x}{\sm}\equiv t = \cosh (\pi b \, \sg).
\eeq
One finally obtains the full cylinder amplitude by integrating over the single
real moduli $\tau$ of the cylinder:
\beq\label{15} 
Z_\m(r,t_1;s,t_2) = \int_0^\infty \d \tau \; Z_{ghost}(q)
Z_{Liouv}(t_1,t_2;q) Z_{mat}(r,s;q).
\eeq
This cylinder amplitude
depends not only on the Cardy states $r,s$, but also on the values
of the boundary cosmological constants $t_1,t_2$ as well as the bulk
cosmological constant $\m$.

From the discussion above it is natural that the matrix model (for a
specific value of $m$) only leads to a single cylinder amplitude
since it corresponds to an explicit (lattice) realization of the
conformal field theory, and thus only to one realization of boundary
conditions. In the language of Cardy states we want to
identify {\it which} boundary condition is realized in the scaling
limits of the one-matrix model. We do that by calculating the
cylinder amplitude \rf{15} and then comparing the result with the
matrix model amplitude.

The calculation, using \rf{9}, \rf{12} and \rf{15}, is in principle
straight forward, but quite tedious, see \cite{ag2} for some
details. The result is for $r+s \leq m$ (for $r+s > m$ we have a
slightly more complicated formula, which we will not present here,
but all conclusions are valid also in this case)\footnote{The prime
in the summation symbol $\sideset{}{'}\sum $ means that the
summation runs in steps of two.} 
\beq\label{16} 
Z_\m(r,t_1;s,t_2) =
- \sideset{}{'}\sum_{k=1-r}^{r-1}\;
\sideset{}{'}\sum_{l=1-s}^{s-1}\log \left(\Big[
(\sqrt{t_1+1}+\sqrt{t_2+1})^2-f_{k,l}(t_1,t_2)\Big]\sm\,a\right)
\eeq 
where $a$ is the cut-off (as in \rf{8a}) and the summations are
in steps of two, indicated by the primes in the summation symbols.
\beq\label{17} 
f_{k,l}(t_1,t_2) = 4 \left[\sqrt{(t_1+1)(t_2+1)} + 2
\cos^2\left(\frac{(k+l)\pi b^2}{4}\right) \right] \; \sin^2 \left(
\frac{(k+l)\pi b^2}{4}\right). 
\eeq 
From eqs.\ \rf{16} and \rf{17}
it follows that {\it we have agreement with the matrix model
amplitude \rf{8a} if and only if $r\equ s\equ 1$}. The $r\equ 1$
boundary condition is in the concrete realizations of conformal
field theories related to the so-called fixed boundary conditions
and for the matter part of the cylinder amplitude it corresponds to
the fact, that only the conformal family of states associated with
the identity operator propagates in the open string channel.

Following Martinec \cite{martinec} it is now possible to calculate
the FZZT--ZZ amplitude by replacing one of the FZZT wave functions
in \rf{12} with 
\beq\label{19} 
\Psi_{\hn}(P)\propto
\Psi_{\sg(\hn)}(P) -\Psi_{\sg(-\hn)}(P), 
\eeq 
where (in the $(2,2m-1$) models) 
\beq\label{19a} 
\sg(\hn) = i\left(\frac{1}{b}+\hn\, b\right), 
\eeq 
and where $ \hn =1, \ldots,
m-1$ is an integer labeling the different principal ZZ-branes.

Notice, the boundary cosmological constants $t_\hn$ and $t_{-\hn}$
corresponding to the complex valued $\sg(\hn)$ and $\sg(-\hn)$ are
real and are actually the same for a given value of $\hn $:
\beq\label{19b} 
t_{\hn} = t_{-\hn} = -\cos \Big(\frac{2\hn\pi}{2m+1}\Big), 
\eeq 
i.e. they are the zeros of the polynomial $P_{m}(t)$ in
formula \rf{1}. We now obtain the following FZZT--ZZ cylinder
amplitude\footnote{The upper sign in \rf{20} is for $0 \leq
k+l+\hn$, while the lower sign is for $k+l+\hn\leq 0$} for $r+s \leq
m$, differentiated after the boundary cosmological constant on the
FZZT brane: 
\bea 
Z_\m'(r,\hn;s,t)&\propto& \sideset{}{'}\sum_{k=-(r-1)}^{r-1}
\;\sideset{}{'}\sum_{l=-(s-1)}^{s-1}\;\;
\frac{(\pm)}{\sm \sqrt{1+t}} \label{20}\\
&&\left[ \frac{1}{\sqrt{t\pl 1} \pl \sqrt{1\pl t_{k + l + \hn}}}-
\frac{1}{\sqrt{t\pl 1} \mi \sqrt{1\pl t_{k+l + \hn}}}\right].
\nonumber 
\eea 
The differentiation after the boundary cosmological
constant is performed in order to compare with the corresponding
amplitude $G_\m(t,t'\equ -t_\hn, d\to \infty)$ given by \rf{6c},
which is the amplitude of a cylinder with one marked point on the
compact boundary.

Let us now consider the FZZT-ZZ cylinder amplitude with an $r \equ
1$ Cardy matter boundary condition imposed on the FZZT boundary.
This is the natural choice if we want to compare with the matrix
model results since the Cardy matter boundary condition captured by
the matrix model is precisely $r \equ 1$. In this case the summation
over $s$ is not present in eq.\ \rf{20} and comparing formula
\rf{20} with the expression \rf{6c} for $G_\m(t,-t_\hn,d \to\infty)$
one can show that
\beq\label{21} 
G_\m(t,-t_\hn,d \to \infty) \propto
\sum_{r=1}^{m-1} S_{1,r} \;Z'_\m(r,\hn;1,t),
\eeq
where $S_{k,l} $
is the modular S-matrix in the $(2,2m\mi 1)$ minimal CFT, i.e.\ \cite{book}
\beq\label{22} 
S_{k,l} = \sqrt{2}\, b \,
(-1)^{m+k+l}\sin (\pi kl\,b^2).
\eeq
This result is valid for any
$(2,2m \mi 1)$ minimal CFT coupled to quantum gravity and is valid
independent of which zero $t_{k}$ the running boundary coupling
constant approaches in the limit $d \to \infty$. The proof of
\rf{21} is straight forward but tedious and will
not be given here (see \cite{ag2} for some details).

The natural interpretation of eq.\ \rf{21} is that the matter
boundary state of the exit loop in the loop--loop amplitude
$G_\m(t,-t_\hn,d)$ is projected on the following linear combination
of Cardy boundary states in the limit $d \to \infty$: 
\beq\label{23}
|a\ra = \sum_{r=1}^{m-1} S_{1,r} \; |r\ra_{\mathrm{Cardy}} \propto
~|1\ra\ra, 
\eeq 
where the last state is the Ishibashi state
corresponding to the identity operator and where we have used the
orthogonality properties of the modular S-matrix and the relation
between Cardy states and Ishibashi states: 
\beq\label{24}
|r\ra_{\mathrm{Cardy}} = \sum_{k=1}^{m-1}
\dfrac{S_{r,k}}{\sqrt{S_{1,k}}} \;|k\ra\ra. 
\eeq

The Ishibashi state corresponding to the identity operator
is in a certain way the
simplest boundary state available,
and it is remarkable that it is precisely this
state which is captured by the explicit transition from compact to non-compact
geometry enforced by taking the distance $d \to \infty$.

\section{The nature of ZZ branes}

A ZZ-brane is defined as the tensorproduct of a ZZ boundary state
$|r,s\ra_{\zz}$ and a Cardy matter state $|k,l\ra_{\cm}$. Hence, in
addition to specifying a ZZ boundary condition, we have to impose a
Cardy matter state at infinity. In the original article by the
Zamolodchikovs only Liouville field theory was considered \cite{zz}.
However, their line of reasoning relied crucially on the
interpretation of quantum Liouville theory as describing 2d quantum
gravity. Invarians under diffeomorphisms demands that the total
central charge is zero. Hence, for a given value of the Liouville
central charge we should think of the corresponding matter and ghost
fields (which have central charges such that the total central
charge is zero) as having been integrated out. In the article of the
Zamolodchikovs the nature of the various boundary states at infinity
was unclear. The successive work in the context of non-critical
string theory \cite{sei1,sei2,martinec2} showed how to reduce the
possible ZZ branes to a number of principal ZZ branes. However, the
origin of precisely these principal ZZ branes remained somewhat of a
mystery.

In $(p,q)$ minimal non-critical string theory the principal
ZZ-branes are defined as
\begin{equation}
|1,1\ra_{\cm} \otimes |r,s\ra_{\zz},
\end{equation}
where $1\leq r \leq p-1$, $1 \leq s \leq q-1$ and $rq-sp>0$. It
turns out that we may interpret the $(p-1)(q-1)/2$ different
principal ZZ branes in $(p,q)$ minimal string theory as matter
dressed basic $(1,1)$ ZZ boundary states \cite{ag3}:
\beq\label{zz1} 
|1,1\ra_{\cm} \otimes |r,s\ra_{\zz} = |r,s\ra_{\cm}
\otimes |1,1\ra_{\zz}. 
\eeq 
Eq.\ \rf{zz1} should be understood in
the following way: With regard to expectation values of physical
observables it does not matter whether we use the right hand side or
the left hand side of eq.\ \rf{zz1}. Thus, in this sense there
exists only one ZZ boundary condition, the basic $(1,1)$ boundary
condition.
Furthermore, we have the following generalization of \rf{zz1}:
\beq\label{zz2} 
|k,l\ra_{\cm} \otimes |r,s\ra_{\zz} =
\left(\sideset{}{'}\sum_{i=|r-k|+1}^{{\rm top}(r,k;p)}
\sideset{}{'}\sum_{j=|s-l|+1}^{{\rm top}(s,l;q)}
|i,j\ra_{\cm}\right) \otimes |1,1\ra_{\zz}, 
\eeq 
where
\beq\label{zz2a} 
{\rm top}(a,b;c) \equiv \min(a+b-1,2c-1-a-b). 
\eeq
Notice, this summation is precisely the same which appears in the
fusion of two primary operators in the $(p,q)$ minimal conformal
field theory: 
\beq\label{zz3} 
O_{k,l} \times O_{r,s} =
\sideset{}{'}\sum_{i=|r-k|+1}^{{\rm top}(r,k;p)}
\sideset{}{'}\sum_{j=|s-l|+1}^{{\rm top}(s,l;q)}  [O_{i,j}]. 
\eeq
Why are eqs.\ \rf{zz1} and \rf{zz2} true? (we refer to \cite{ag3}
for the full details of the proof.)

Recall the definition of the Cardy matter boundary states in the
$(p,q)$ minimal conformal field theory: 
\beq\label{zz3a} 
|k,l\ra_\cm \equiv \sum_{i,j} 
\frac{S(k,l;i,j)}{\sqrt{S(1,1;i,j)}}\; \vv i,j\rara 
\eeq 
where the summation runs over all the different
Ishibashi states $\vv i,j\rara$ in the $(p,q)$ minimal model and
\beq\label{zz3b} 
S(k,l;i,j) = 2\sqrt{\frac{2}{pq}}
(-1)^{1+kj+li}\sin (\pi b^2 lj)\;\sin(\pi k i /b^2), 
\eeq 
is the modular S-matrix in the $(p,q)$ minimal model. The Cardy matter
boundary states are labeled by two integers $(k,l)$, which satisfy
that $1 \leq k \leq p\mi 1$, $1\leq l \leq q\mi 1$ and $ kq-lp > 0$.

On the other hand the principal ZZ boundary states are defined as
\beq\label{zz3c} 
|r ,s\ra_{\zz} = \int_0^\infty \d P \; \frac{\sinh
( 2\pi r P/b) \; \sinh (2\pi s P b)}{\sinh ( 2\pi P/b) \; \sinh (
2\pi P b)}  \; \Psi_{1,1} (P) \, \vv P \rara, 
\eeq 
where $b \equ\sqrt{p/q}$. $\Psi_{1,1}(P)$ is the basic ZZ wave function
\cite{zz}: 
\beq\label{zz3d} 
\Psi_{1,1}(P) = \b \frac{iP
\m^{-iP/b}}{\G(1-2iPb)\G(1-2iP/b)}, 
\eeq 
where the constant $\b$ is
independent of the cosmological constant $\m$ and $P$. Finally, $\vv
P\rara$ denotes the Ishibashi state corresponding to the non-local
primary operator $\exp(2(Q/2+iP)\phi)$ in Liouville theory, where
$Q\equ b \pl 1/b$.

Notice, the ranges of the indices $k,l$ labeling the different Cardy
matter boundary states and the indices $r,s$ labeling the principal
ZZ branes are the same. As noted already by the Zamolodchikovs in
\cite{zz}, the modular bootstrap equations for the ZZ boundary
states are surprisingly similar to the bootstrap equations for the
Cardy matter boundary states in the minimal models. The key point is
now that the physical operators in minimal string theory carry both
a matter ``momentum'' and a Liouville ``momentum'' and these are not
independent, but related by the requirement that the operators scale
in a specific way. In particular, the Liouville momenta $P$ of the
physical observables are imaginary and the imaginary $i$ explains
the shift from $\sin$ to $\sinh$ going from \rf{zz3a} to \rf{zz3c}.
The coupling between the matter and Liouville momenta implies, that
physical expectation values will be the same irrespectively of
whether we use the left or the right side of eq.\ \rf{zz1}.

Our interpretation of the term
\begin{equation}\label{dres}
\frac{\sinh ( 2\pi r P/b) \; \sinh (2\pi s P b)}{\sinh ( 2\pi P/b)
\; \sinh ( 2\pi P b)}
\end{equation}
in the definition of the principal $(r,s)$ ZZ boundary state
\rf{zz3c} as a dressing factor arising from the integration over the
matter and the ghost fields becomes evident when considering the
cylinder amplitude. For simplicity we only consider this amplitude
in $(2,2m-1)$ minimal string theory. The cylinder amplitude does not
factorize into a matter part and a Liouville part. The integration
over the single real moduli $\tau$ correlates matter with geometry.
If one imposes the $(1,s)$ Cardy matter state on a ZZ boundary and
performs the integrations over both $\tau$, the matter and the ghost
fields, the ZZ boundary wave function get dressed exactly with the
term \rf{dres} with $r=1$. \cite{ag3}

\section{Discussion}

We have shown how it is possible to construct an explicit transition
from compact to non-compact geometry in the framework of 2d quantum
gravity coupled to conformal field theories. The non-compact
geometry is AdS-like in the sense that the average area and the
average length of the exit loop diverge exponentially with $d$ when
$d \to \infty$ as shown in \cite{aagk} (for pure gravity), and the
corresponding amplitude can be related to the FZZT-ZZ cylinder
amplitude with the simplest Ishibashi state living on the ZZ brane.
The $d \to \infty$ limit plays an instrumental role and we would
like to address two important aspects of this.

Firstly, our construction also adds to the understanding of the
relation \rf{19} discovered by Martinec. In Liouville theory there
is a one-to-one correspondance between the ZZ boundary states
labeled by $(m,n)$ and the degenerate primary operators $V_{m,n}$
\cite{zz}. This correspondance completely determines the Liouville
cylinder amplitude with two ZZ boundary conditions: The spectrum of
states flowing in the open string channel between two ZZ boundary
states is obtained from the fusion algebra of the corresponding
degenerate operators. Similarly, there is a one-to-one
correspondance between the FZZT boundary states labeled by $\sg>0$
and the non-local "normalizable" primary operators
$V_{\sg}=\exp((Q+i\sg)\phi)$, where $\phi$ is the Liouville field.
The conformal dimension of the spin-less degenerate primary operator
$V_{m,n}$ is given by 
\beq\label{29}
\Delta_{m,n}=\frac{Q^{2}-(m/b+nb)^{2}}{4}, 
\eeq 
while the conformal
dimension of the spin-less non-local primary operator $V_{\sg}$ is
given by 
\beq\label{30} 
\Delta_{\sg}=\frac{Q^{2}+\sg^{2}}{4}. 
\eeq
Since $\Delta_{m,n}=\Delta_{\sg}$ for $\sg=i(m/b+nb)$, one is
naively led to the wrong conclusion, that a FZZT boundary state
turns into a ZZ boundary state, if one tunes $\sg=i(m/b+nb)$.
However, the operator $V_{m,n}$ is degenerate and in addition to
setting $\sg=i(m/b+nb)$ we therefore have to truncate the spectrum
of open string states, that couple to the FZZT boundary state, in
order to obtain a ZZ boundary state. This is precisely captured in
the relation \rf{19} concerning the principal ZZ boundary states.
The world-sheet geometry characterizing the FZZT brane is compact,
while the world-sheet geometry of the ZZ-brane is non-compact.
Hence, truncating the spectrum of open string states induces a
transition from compact to non-compact geometry.

In order to clarify how this truncation is obtained in our concrete
realization of a transition from compact to non-compact geometry, we
have to discuss the boundary cosmological constant of the exit loop.
The cylinder amplitude \rf{6} may be expressed as
\begin{equation}
G_{\mu}(x,y;d) = \int_{0}^{\infty}\,dl e^{-yl} G_{\mu}(x,l;d)
\end{equation}
where the cylinder amplitude $G_{\mu}(x,l;d)$ with fixed length $l$
of the exit loop is given by
\begin{equation}
G_{\mu}(x,l;d) = e^{-\bx(d)l} \;\frac{w_\m(\bx(d))}{w_\m(x)}
\end{equation}
Hence, an interpretation of the running boundary coupling constant
$\bt=\bx(d)/\sqrt{\mu}$ (measured in units of $\sqrt{\mu}$) as a
boundary cosmological constant \emph{induced} on the exit loop seems
obvious. Notice, this induced boundary cosmological constant
approaches one of the values $t_{k}$ associated with the ZZ-branes
in the limit $d \to \infty$. However, an AdS geometry emerges in the
limit $d \to \infty$ if and only if we set the boundary cosmological
constant of the exit loop $y/\sqrt{\mu}=-t_{k}$.

The induced boundary cosmological constant approaches one of the
zeros $t_{k}$ in the limit $d \to \infty$ regardless of whether we
set $y/\sqrt{\mu}=-t_{k}$ i.e. regardless of whether we generate an
AdS geometry or not. Hence, these discrete values of the boundary
cosmological constant induced at infinity seems to be generic to
non-compact geometries. This suggests, that we should regard the
boundary cosmological constants associated with the ZZ-branes as
induced.

Secondly, in \cite{sei1} it was advocated that the algebraic surface
\beq\label{25} T_p(w/C_{p,q}(\mu)) =T_q(t), \eeq where $C_{p,q}(\m)$
is a constant, is the natural "target space" of $(p,q)$ non-critical
string theory. For $(p,q)=(2,2m \mi 1)$ eq.\ \rf{25} reads
\beq\label{25a} w^2 = \mu^{\frac{2m-1}{2}}P_m^2(t)(t+1), \eeq and in
this case the extended target space is a double sheeted cover of the
complex $t$-plane except at the singular points, which are precisely
the points $(t_k,w\equ 0)$ associated with the zeros of the
polynomial $P_m(t)$. One is also led to this extended target space
from the world-sheet considerations made here. We want the running
boundary coupling constant to be able to approach any of the fixed
points $t_k$ in the limit $d \to \infty$, i.e.\ we want all the
fixed points to be attractive. This is only possible if we consider
the running boundary coupling constant $\bt(d)= \bx(d)/\sm$ as a
function taking values on the algebraic surface defined by \rf{25a}.
The reason is that $t_k$ is either an attractive or a repulsive
fixed point depending on which sheet we consider and some of the
fixed points are attractive on one sheet, while the other fixed
points are attractive on the other sheet. Hence, we are forced to
view $\bt(d)$ as a map to the double sheeted Riemann surface defined
by eq.\ \rf{25a} in the $(2,2m \mi 1)$ minimal model coupled to
quantum gravity.

The picture becomes particularly transparent if we use the
uniformization variable $z$ introduced for the $(p,q)$ non-critical
string in \cite{sei1} by 
\beq\label{26}
t=T_p(z),~~w/C_{p,q}(\m)=T_q(z), 
\eeq 
i.e.\ in the case of $(p,q)=(2,2m \mi 1)$: 
\beq\label{26a} 
z=\frac{1}{\sqrt{2}} \sqrt{t+1}. 
\eeq 
The map \rf{26} is one-to-one from the complex
plane to the algebraic surface \rf{25}, except at the singular
points of the surface where it is two-to-one. The singular points
are precisely the points corresponding to ZZ branes. If we change
variables from $x$ to $z$ in eq.\ \rf{5} (choosing $\m \equ 1$ for
simplicity) we obtain 
\beq\label{27}
\frac{\prt }{\prt d}\; \tG_\m (z,z';d) = - \frac{\prt }{\prt z} \;
\tP_m(z) \tG_\m (z,z';d), 
\eeq 
where $\tG_\m(z,z';d)=zG_\m(x,y;d)$
and where the polynomial $\tP_{m}(z)$ is 
\beq\label{28} 
\tP_{m}(z) \propto \prod_{k=1}^{m-1} (z^2 -z_k^2), ~~~~z_k = \sin
\left(\frac{\pi}{2} \, b^2 \, k\right). 
\eeq 
Each zero $t_k$ of
$P_m(t)$ gives rise to two zeros $\pm z_k$ of $\tP_m(z)$. The zeros
$\pm z_{k}$ are the fixed points of the running ``uniformized''
boundary cosmological constant $\bar{z}$ associated with the
characteristic equation corresponding to eq.\ \rf{27}. For a given
value of $k$ one of the two zeros $\pm z_k$ is an attractive fixed
point, while the other is repulsive. Moving from one sheet to the
other sheet on the algebraic surface (\ref{25a}) corresponds to
crossing the imaginary axis in the $z$-plane. Hence, for a given
value of $k$ the two fixed points $\pm z_{k}$ are each associated
with a separate sheet and $\bar{z}$ will only approach the
attractive of the two fixed points $\pm z_{k}$, if $\bt(d)$ belongs
to the correct sheet.

Quite remarkable eq.\ \rf{27} was derived in the case of pure 2d
gravity (the $(2,3)$ model corresponding to $c\equ 0$) using a
completely different approach to quantum gravity called
CDT\footnote{It should be noted that CDT seemingly has an
interesting generalization to higher dimensional quantum gravity
theories \cite{ajl1}} (causal dynamical triangulations) \cite{al}
and the uniformization transformation relating the CDT boundary
cosmological constant $z$ to the boundary cosmological constant $t$
was derived and given a world-sheet interpretation in \cite{ackl},
but again from a different perspective. From the CDT loop-loop
amplitude determined by \rf{27} one can define a CDT ``ZZ brane''
with non-compact geometry \cite{ajwz}.

\end{document}